\documentclass[prd,aps,twocolumn,showpacs,preprintnumbers,%
amsmath,amssymb]{revtex4}

\usepackage{graphicx}   

\def\Pslash{\hbox{/\kern-.65em$P$}}
\def\Kslash{\hbox{/\kern-.65em$K$}}

\begin{document}

\title{Determination of the age of the earth from Kamland
 measurement of geo-neutrinos }

\author{ Subhendra Mohanty}
\affiliation{ Physical Research Laboratory, Ahmedabad 380009,
India.\\ }

\date{\today }

\begin{abstract}
The low energy component of the anti-neutrino spectrum observed in
the recent Kamland experiment has significant contribution from
the radioactive decay of $^{238}U$ and $^{232}Th$ in the earth.
 By taking the ratio of the anti-neutrino
events observed in  two different energy ranges we can determine
the present  value of the Thorium by Uranium abundance ratio, independent
of the U,Th distribution
in the earth. Comparing the present abundance ratio  with the r-process
predicted initial value 
 we determine the age of the earth as a function of
$\Delta m^2$ and $Sin^2 2 \theta$. We find that the age of the
earth determined from KamLAND data matches the age of solar system
($4.5 Gyrs$ determined from meteorites) for the LMA-I solution. For the
 LMA-II
solution  the age of the earth
does not match the solar system age even at $90 \% C.L.$

 \end{abstract}

\maketitle

\begin{figure}
 \label{Fig.1}
\centering
\includegraphics[width=7cm]{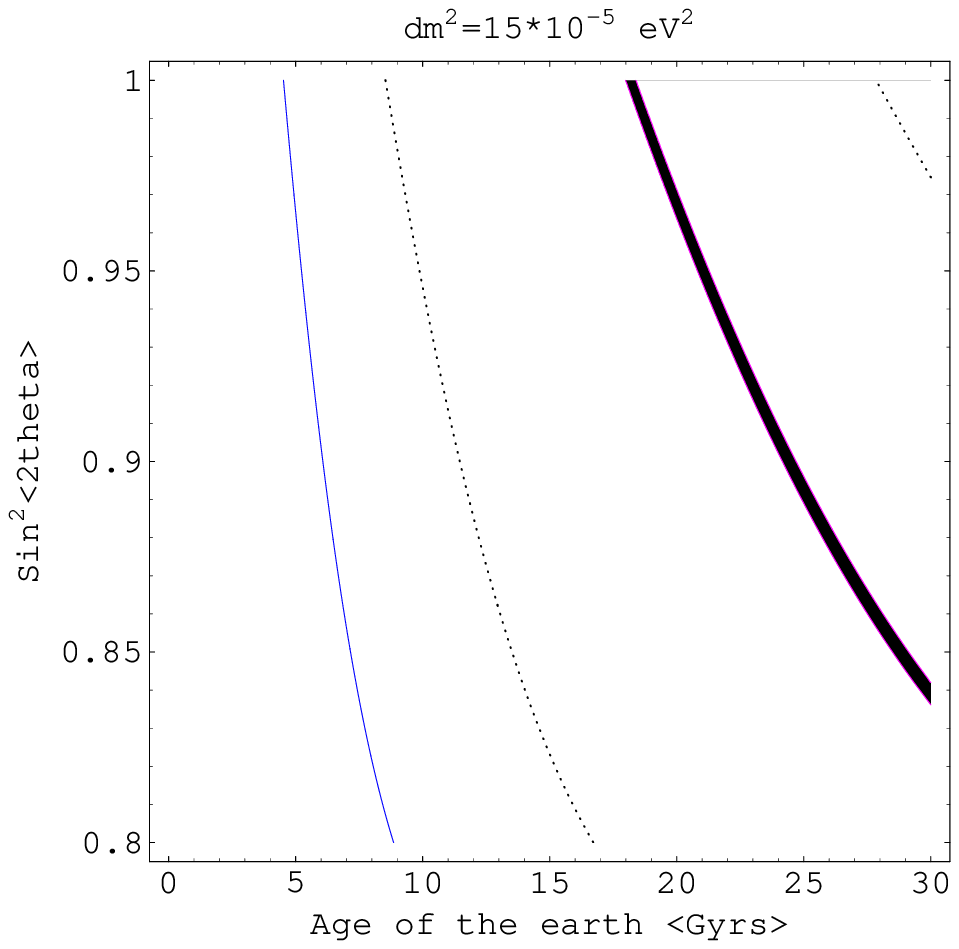}
\includegraphics[width=7cm]{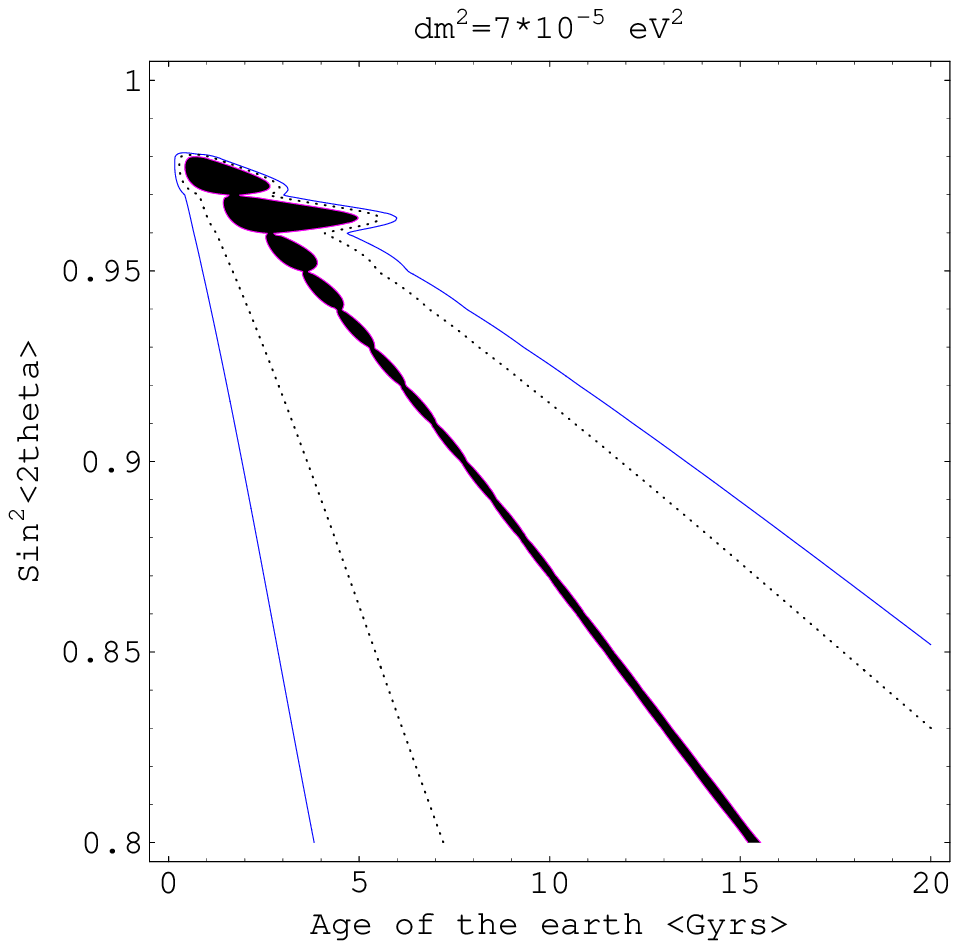}
\caption{ Age of the earth at $68.3 C.L. \%$ (dashed curves) and $90\%
C.L.$ (solid curves). Shaded region is the $\chi^2 \simeq 0$ allowed
parameter
space. }
\end{figure}

The recent results from  KamLAND  \cite{kam} are
significant for establishing the LMA solution of the solar
neutrino problem. In addition they present the  first
statistically significant  measurement the anti-neutrinos from
radioactive decay of $^{238}U$ and $^{232}Th$ present in the crust
and mantle of the earth. The possibility of detecting geo
neutrinos from radioactivity in the the earth was first raised
by Eder \cite{eder} and was revived by Krauss et al \cite{krauss}
and Kobayashi and Fukao \cite{koba}. Quantitative spectra of the
anti-neutrino events which could be observed at Kamioka and Gran
Sasso  from radioactivity of $Th$ and $U$ in the earth has been given by
Rothschild et al \cite{roth} and  Raghavan et al \cite{ragh}. The
emphasis of these papers and the recent work of Fiorentini et al
\cite{fior1} is to use the observations of $\bar \nu_e$ events at
KamLAND and Borexino experiments to determine the distribution of
$U, Th$ in the crust and mantle of the earth, and determine what
fraction of the $40 TW$ heat generated by earth from
radioactivity. KamLAND \cite{kam} reports  observing $9$ $\bar
\nu_e$ events below $2.6 MeV$ (visible positron energy) which  are
ascribed to $U,Th$ radioactivity in the earth. Subsequently a
 analysis of the geo neutrinos from KamLAND observations
has been done by Fiorentini et al \cite{fior2} with the aim of determining
the geological distribution of $U$ and $Th$ in the earth. Fiorentini et
al. assume that $[Th/U]=3.8$ the solar system 
value and try to determine the $U$ and $Th$ content of the core and
mantle. 

In this paper we have a different aim which  is to extract the average
thorium by uranium
abundance ratio, $[Th/U]$ from the KamLAND data, as this parameter is
an ideal chronometer
for measurement of cosmological time-scales. $Th$ and $U$ present in the
earth 
are produced in
supernova by
r-process nucleosynthesis \cite{arnett}. The theoretical prediction 
initial abundance ratio is
robust against perturbations of the astrophysical parameters at the site of
the r-process to 
within  $\pm 5\%$ \cite{otsuki}. By determining the present abundance
ratio of $[Th/U]$ in the earth from KamLAND, we can measure the age of
the earth-i.e the time elapsed between the supernova explosion where the
 $Th$ and $U$ in the earth were produced, and the present. KamLAND
observes
geo anti-neutrinos in the (positron signal) energy range $0.9 MeV < E <
2.5 MeV$.
Thorium decay neutrinos have a maximum $E$ value of $1.5 MeV$ whereas
Uranium neutrinos
contribute in the entire observed energy range of the geo signal. The
geo neutrinos undergo an energy independent suppression as their average
distance from the detector $\sim 10^3 km$ is much larger than the
oscillation length $\sim 10^2 km$. We take the ratio of the geo events,
in the the $(0.9 -1.5) MeV$ bins and the geo events, in the
$(1.5-2.5) MeV$
energy bins to determine the $[Th]/[U]$. This enables us to
determine the age of the earth independent of the geological
 distribution of $Th$ and $U$ in the earth.

 Using the KamLAND data we can determine the
age of the earth as a function of $\Delta m^2$ and $Sin^2 2 \theta$, as
these parameters determine the reactor background that has to be
subtracted
from the KamLAND observed events to arrive at the geo-neutrino signal.
In Fig.1 we show the age of the earth as a function of the mixing angle
for the LMA-I ($\Delta m^2 =7 \times 10^{-5} eV^2$) and LMA-II
($\Delta m^2 =15 \times 10^{-5} eV^2$ ) solutions \cite{kamfit}. We
estimate the
total uncertainity in the age (for a given $\Delta m^2, Sin^2
2\theta$) to be $\pm 36.5\%$. The central shaded regions are the best fit
points ($\chi^2 \sim 0$) and the two outer lines enclose the allowed
region of parameter space with $C.L.$ of $68.3\%$ and $90 \%$
respectively.  We see from Fig.1 that
for the LMA-II solution $t_{age} > 5.0 Gyrs$ ( at $90\% C.L$) and
$t_{age}
>9.0 Gyrs $ (at $68.3 \% C.L$) . It is known
from
the radiochemical dating of meteorites that the age of the oldest
meteorites (the chondrites) is $4.5 Gyrs$ \cite{andres}. If the age of the
earth is  close to the age of the chondrites then the LMA-II solution 
is ruled out. We see from Fig1 that for the LMA-I solution, the
age of the earth agrees with the age of the meteorites for the mixing
angle in the range $0.88 < Sin^2
2 \theta < 0.93$. 

The KamLAND detector consists of about $1 kiloton$ of liquid
scintillator  surrounded by photo-multiplier tubes.
 Electron anti-neutrinos are detected by means of the
inverse beta decay $ \bar \nu_e +p \rightarrow e^+ +n  $ by
looking for the $e^+$ in delayed coincidence with the $2.2 MeV $
$\gamma$-ray from the neutron capture by protons ($n+p \rightarrow
d + \gamma$). The $e^+$ annihilate in the detector producing a
total visible energy $E$ which is related to the initial $\bar
\nu_e$ energy,$E_\nu$, as $E=E_\nu -(m_n-m_p+m_e) + 2 m_e=E_\nu -
0.78 MeV$. Only those radioactive product $\bar \nu_e$'s with
energies above the inverse beta decay reaction threshold of
$m_n-m_p+m_e=1.8 MeV$ can be detected. The main source of
 $\bar \nu_e$'s and radiogenic heat in the earth are decays of
$^{238}U$, $ ^{232}Th$ and $^{40}K$. The $\bar \nu_e$ from
$^{40}K$ decay have $E_{\nu max}=1.31~ MeV$ and will not register
in the KamLAND detector. In the decay chain of $^{238}U$,
only $\bar \nu_e$'s from the $\beta$ decays of $^{234}Pa~ $
and $^{214}Bi~ $ are above the threshold for
detection in KamLAND. In the $^{232}Th$ decay chain, $\bar
\nu_e$'s from the beta decays of $^{228}Ac~$
$^{212}Bi $  contribute to the KamLAND signal.
 Thorium decay $\bar \nu_e$'s will contribute only
to the (positron signal) energy bins below $1.5~ MeV$ whereas
Uranium $\bar \nu_e$'s will contribute to all energy bins below
$E=2.5 MeV$. This fact enables us to separate the thorium neutrino
signal from that of uranium. The energy spectrum of $\bar \nu_e$'s
from each of these beta decays can be expressed analytically as
follows,
 \begin{eqnarray}
\eta_X(E)&=& \sum A(Q,Z) \times F(Z,E_\nu) \nonumber \\ &\times&
E_\nu^2 (Q+m_e -E_\nu)
 [(Q+m_e-E_\nu)^2-m_e^2)]^{1/2} \nonumber \\
 \label{spectrum}
 \end{eqnarray}
where $F(Z,E_\nu)$ is the Fermi function that accounts for the
distortion of the spectrum due to Coulomb attraction of the
outgoing $e^-$ with the nucleus, the sum is over each of the beta
decays in the $X~ (=Th, U)$ decay chain with $Q$ value above the $1.8
MeV$ threshold,  $A(Q,Z)$ are constants obtained by the
normalizing the spectrum for each term in the sum to unity. The
remaining terms are kinematical factors for the two body decay
(assuming the recoil energy of the nucleus is negligible). Other
nuclear physics effects can be parameterized by adjusting the
overall normalization to match the tabulated experimental values
\cite{Behrens} ( we fit the normalization of $U$ and $Th$ spectrum
by requiring that  $I=\int dE \sigma(E) \eta(E)= 0.51~ (2.52)
\times 10^{-44} cm^2$ for Thorium (Uranium) where $\sigma$ is the
inverse beta decay cross section shown in
(\ref{sigma}) below.

 The neutrino flux from the earth at a location $\vec
R_d$ of the detector can be expressed as the integral
\begin{eqnarray}
\Phi_\nu(\vec R_d)=\frac{1}{4 \pi}\int d^3 r \frac{1}{|\vec r-\vec
R_d|^2 }\frac{n_X(\vec r)}{\tau_X} P_{ee}(|\vec r-\vec R_d|)
\end{eqnarray}
where $n_X(\vec r)$ is the number density of the radioactive atoms
$X ~(=U,Th)$ and $\tau_X$ is the lifetime of $X$ and $Pee(|\vec
r-\vec R_d|)$ is the $\bar \nu_e$ survival probability. Assuming
that $n_X(\vec r)$ is approximately constant within a spherical shell, the
expression for the flux from a shell of constant density of
radioactive atoms can then be written as
\begin{eqnarray}
\Phi_\nu=\frac{G_i }{4 \pi~ R_e^2~} \frac{M_{i}~ [X]_i}{ \tau_X}
\label{phi}
 \end{eqnarray}
where  $[X]_i$ is the number of $X$ atoms per unit mass  in the
shell $i$ ,  $M_{i}$ is the mass of the $i$' th shell and $ R_e$
is the radius of the earth.
 The geometrical factor $G_i$ depends upon the thickness of the
 shell
 at the site of the detector and
is given by
\begin{eqnarray}
G_i= \frac{3}{2}\frac{1}{ (x_2^3 -x_1^3)}\int_{x_1}^{x_2} dx
\int_{-1}^1 d\mu \frac{x^2}{1+x^2-2 \mu x}~ P_{ee}
\end{eqnarray}
where $x_1=r_1/R_e, x_2=r_2/R_e$ are the inner and outer radii of
the shell in units of $R_e$, $\mu$ is the cosine of the angle
between the position of the detector, $\vec R_d$, and a point
$\vec r$ inside the shell (the origin of the coordinates is chosen
at the center of the earth). The survival probability of $\bar
\nu_e$ is a function of the distance between the detector and the
point $\vec x$ in the shell and is given explicitly by
\begin{eqnarray}
P_{ee}= 1-sin^2 2\theta ~sin^2 \left[\frac {\Delta m^2 R_{e}}{4
E_\nu}(1+x^2 -2 \mu x)^{1/2} \right]
\end{eqnarray}
Inserting $P_{ee}$ in the expression for the geometrical factor
and carrying out the integration over the shell thickness $x$ and
the angular variable $\mu$, we can express the energy dependence
of $G_i$ as
\begin{equation}
G_i(E)= G_i(0)(1-\frac{1}{2} ~Sin^2 2\theta \times f_i(E))
\end{equation}
where $G_i(0)$ depends on the shell thickness and $f_i(E)$ is
slowly varying function of $E$ (when $\Delta m^2\simeq
10^{-4}-10^{-5} eV^2$). For the continental crust,
with thickness $30 km$, $G_{cc}(0)= 3.54$ and $f_{cc}(E)$ is a
monotonically decreasing function of the positron signal energy,
$E$, with $f_{cc}(0.9)=1.00$ and $f_{cc}(2.5)=0.98$. For the
oceanic crust with thickness of about $6km$, $G_{oc}=4.34$ and
$f_{oc}(E)$ decreases monotonically from $f_{oc}(0.9)=0.88$ to
$f_{oc}(2.5)=0.82$.  For neutrinos coming from
the mantle, there is an energy independent
suppression, and we have $G_m(0)=1.5$ with $f_m(E)=1$.

The total contribution to the neutrino flux  from the crust and
mantle can be written as
\begin{eqnarray}\label{flux}
  \Phi_\nu = \left(\frac{\eta_U(E)}{\tau_U} +
  \left[\frac{Th}{U}\right]~\frac{\eta_{Th}(E)}{
  \tau_{Th}}\right)
  \left[{\displaystyle \sum_{i}}
    M_i~[U]_i ~ g_i~G_i (E)\right]\nonumber\\
\end{eqnarray}
where the index $i=cc,oc,m$ denotes the continental crust,oceanic
crust and mantle respectively.
In the absence of significant chemical segregation of $U$ and $Th$
the ratio $([Th]_i/[U]_i)$ can be taken to be the  same in the
crust and the mantle. 
 $g_i$ is the location parameter which represents
the fraction of the the continental crust {\it vis-a-vis} the
oceanic crust surrounding the detector . For a detector in Japan
which has the oceanic crust on one side and the Asian continental
crust on the other side we may take $g_{cc}\sim g_{oc} \sim 0.5$.

 The
number of detection events $N_i$ in an energy bin centered at $i$
is,
\begin{equation}
N_i= (n_p~ t~ d_{eff})\int_{i-\epsilon/2}^{i+\epsilon/2} dE
~\sigma(E) ~\eta_X(E)~\Phi_\nu
\end{equation}
where $n_p$ number of free target protons in the fiducial volume
($3.46 \times 10^{31}$ for this experiment) , t is the exposure
time ($145.1 days$) and $d_{eff}$ is the detector efficiency
($78.3 \%$) and
 $\epsilon$ is the width of the energy bins ($0.425 MeV$). The low energy 
cutoff in  
KamLAND  is $0.9 MeV$.
 
 The
cross section for the $\bar \nu_e +p \rightarrow e^+ + n$ reaction
is given by \cite{vogel1}
\begin{equation}
\sigma=0.0952 (\frac{E_e~ p_e}{MeV^2}) \times 10^{-42} cm^2
\label{sigma}
\end{equation}
where $E_e=E_\nu-(m_n-m_p)$ is the positron energy and $p_e$ is
the corresponding momentum. 
 Thorium neutrinos have a $ E <1.5 MeV$, whereas Uranium neutrinos contribute in the entire
range of $(E_{min}-2.5) MeV$ (where $E_{min}=0.9$ is the threshold
of the lowest energy bin). The ratio of the neutrino events in the
energy bins between $(E_{min}-1.5) MeV$, $N_I$, and the events in
energy bins $(1.5-2.5) MeV $ , $N_{II}$, depends only on the Thorium
to Uranium ratio $[Th]/[U]$, and the spectral shape of $U$ and $Th$
neutrinos folded with the cross section. The geology factor in the
square brackets in (\ref{flux}) cancels out as it is independent of
energy (to $\pm 1 \%$). Specifically the ratio of the geo-neutrino
 events in the two energy ranges can be written in a simple form
\begin{equation}\label{N12}
  \frac{N_{I}}{N_{II}}= \alpha  +  \left[\frac{Th}{U}\right] \times  \beta
\end{equation}
where
\begin{equation}\label{alpha}
  \alpha=\frac {\int_{E_{min}}^{1.5} dE ~\sigma(E)~ \eta_U(E)}
  {\int_{1.5}^{2.5} dE ~\sigma(E)~ \eta_U(E)}=0.89
\end{equation}
and
\begin{equation}\label{beta}
  \beta= \frac{\tau_U}{\tau_{Th}} \frac {\int_{E_{min}}^{1.5}~ dE ~\sigma(E) ~\eta_{Th}(E)}
  {\int_{1.5}^{2.5} dE ~\sigma(E)~\eta_U(E)}=0.12 ~~~~.
\end{equation}

One can determine the ratio $N_I/N_{II}$  and from that determine
the global average abundance ratio of Th by U using \ref{N12}.
From the initial r-process abundance \cite{otsuki}
 $[Th/U]_0=1.169 \pm 0.08 (1 \sigma)$ and the decay
lifetimes  of $U$ and $Th$ ($\tau_U=6.45 Gyrs, \tau_{Th}=20.03 Gyrs$) we
can relate the time elapsed (in Giga-years) 
between
between the supernova explosion where the earths $U,Th$  were
formed and the present,
\begin{equation}
[Th/U]=[Th/U]_0~~ \exp
[~t_{age}~(\frac{1}{6.45}-\frac{1}{20.03})]~~~.
\end{equation}
We can directly relate the age of the earth to the experimentally measured
quantity $N_I/N_{II}$ as
\begin{equation}\label{age}
  t_{age}= 9.45 \ln \left[\frac{1}{\beta~
  [Th/U]_0}\left(\frac{N_I}{N_{II}}-\alpha \right) \right]~~Gyrs
\end{equation}

We calculate the reactor background using the procedure given in
\cite{mur}. We sum the neutrino flux from  $16$  reactors 
using their power production and distances as input and assuming
an average fuel composition in the ratio $ ^{235}U : ^{238}U :
^{239}Pu : ^{241}Pu :: 0.568 : 0.078 : 0.297 : 0.057$. The reactor
neutrino flux is shown in Fig 2. We see that there although there
is a significant difference between the reactor flux for the LMA-I
 and LMA-II  solutions \cite{kamfit} in the geo-neutrinos energy regime.

The Kamland experiment reports the events in $(0.9-1.75) MeV$ and
$(1.75-2.6) MeV$ energy. Using the theoretical spectrum of the $Th$ and
$U$ geo-neutrinos  we have distributed the $67 \%$ of the events in the 
$(0.9-1.75) MeV$ bin to $(0.9-1.5) MeV$ and $33 \% $ of this bin have been
assigned to the $(1.5-2.5) MeV$ events.
 In the $(0.9-1.5) MeV $ energy range LMA-II reactor neutrinos are more
suppressed than LMA-I, while in the $(1.5 -2.5) MeV$ energy range
the LMA-I reactor neutrinos are more suppressed than LMA-II. 
 After this reactor signal is subtracted from
the KamLAND observed events, the ratio of geo events in the two
energy windows , $N_I/N_{II}$ will go up as we go from from LMA-I to
LMA-II. 

\begin{figure}
 \label{Fig.2}
\centering
\includegraphics[width=7cm]{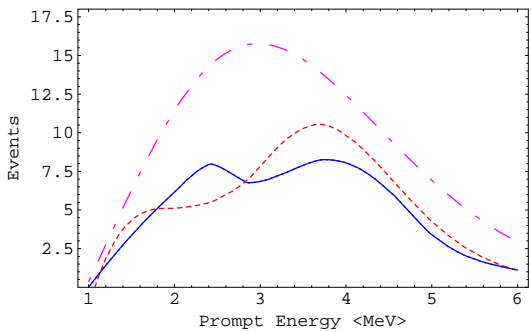}
\caption{Reactor neutrino flux. Dash-dot no oscillation, dotted-
LMA-I, continuous LMA-II. }
\end{figure}

\begin{figure}
 \label{Fig.3}
\centering
\includegraphics[width=7cm]{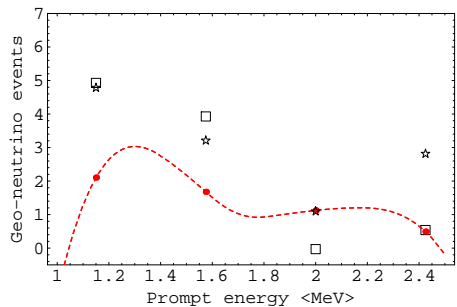}
\caption{Simulated geo-neutrino events shown connected with dotted
lines. Stars (LMA-I) and squares (LMA-II) denote observed events at
KamLAND
minus the reactor background.}
\end{figure}

 In Fig 3. we plot the 
simulated geo-neutrino events shown by points connected by dotted
line - where we have used the input values of $[Th/U]=3.8, Sin^2
2\theta=0.88$, $ M_{cc}(U)= 4.2 \times 10^{17}kg$, 
 $M_{oc}(U)=4.8\times 10^{15} kg$    and $M_{m}(U)=1.23 \times 10^{17}
kg$. The geo-neutrino events have no dependence on $\Delta m^2$. In the
same figure we have show the experimental points (Kamland events minus the 
reactor
background) where stars represent LMA-I and squares represent LMA-II.
Both the LMA-I and LMA-II data points agree with the geo-signal within the
$1 \sigma$ error bars
of the KamLAND data points.The $N_{I}/N_{II}$ ratio which can be read off from the
graph is considerably large for the LMA-II points compared to the LMA-I
points which leads to a large value of $t_{age}$ for LMA-II compared to
LMA-I. The error in determination of $t_{age}$ from KamLAND is estimated as
follows. The systematic error in each of $N_I$ and $N_{II}$  from Table II
of
\cite{kam} is $6.4 \%$ . The combined systematic and statistical error in
$N_I$ is $25 \%$
 and in $N_{II}$ it is $26 \%$. The total systematic and statistical error
in determination of
$N_I/N_{II}$ is $36 \%$. The theoretical uncertainity in the initial
r-process prediction of $[Th/U]$ is $5\%$. The assumption that the
geometrical factor $G_i(E)$ is energy independent introduces an error of
$1\%$.
  Added in quadrature the total
error in
$t_{age}$ is turns out to be $36.5 \%$. With this error we plot the $90
\%$ and $68.3 \%$
allowed region for $t_{age}$ as a function of mixing angle in Fig 1. 
We see that the LMA-II solution does not overlap with the solar system age 
of $4.5 Gyrs$ at $90 \% C.L.$. We emphasize that the LMA-II data points
fit the geo events in KamLAND, its the extra constraint of requiring
$N_{I}/N_{II}\sim 1.13$ (which amounts to $t_{age}=4.5 Gyrs$) which the
LMA-II solution does not fulfill. We must add the cautionary caveat that
although there is a large magnification in the ratio $N_I/N_{II}$ in
going from LMA-I to LMA-II, this ratio is meaningful when $N_I$ and
$N_{II}$ are non-zero. The present KamLAND observations \cite{kam} of
geo-neutrino events are
consistent
with zero at $2 \sigma$, so the conclusions derived in this paper should
be
treated as $1 \sigma$ results.  

{ \bf Acknowledgments} I thank Anjan Joshipura for his help and for
valuable discussions at every stage of this work.

\end{document}